\documentclass[prb,twocolumn,amsmath,amssymb,showpacs]{revtex4}

\usepackage{graphicx}
\usepackage{dcolumn}
\usepackage{bm}

\begin{document}


\title{Hyperfine interaction induced decoherence of electron spins in
quantum dots}

\author{Wenxian Zhang,$^1$ V. V. Dobrovitski,$^1$ K. A. Al-Hassanieh,$^{2,3}$
E. Dagotto,$^{2,3}$ and B. N. Harmon$^1$}

\affiliation{$^1$Ames Laboratory, Iowa State University, Ames, Iowa 50011, USA}

\affiliation{$^2$Department of Physics, University of Tennessee,
Knoxville, TN 37831, USA}

\affiliation{$^3$Condensed Matter Science Division, Oak Ridge
National Laboratory, Oak Ridge, TN 37996, USA}

\date{\today}

\begin{abstract}
We investigate in detail, using both analytical and numerical
tools, the decoherence of electron spins in quantum dots (QDs)
coupled to a bath of nuclear spins in magnetic fields or with
various initial bath polarizations, focusing on the longitudinal
relaxation in low and moderate field/polarization regimes. An
increase of the initial polarization of nuclear spin bath has the
same effect on the decoherence process as an increase of the
external magnetic field, namely, the decoherence dynamics changes
from smooth decay to damped oscillations. This change can be
observed experimentally for a single QD and for a double-QD setup.
Our results indicate that substantial increase of the decoherence
time requires very large bath polarizations, and the use of other
methods (dynamical decoupling or control of the nuclear spins
distribution) may be more practical for suppressing decoherence of
QD-based qubits.
\end{abstract}


\maketitle

\section{Introduction}

Quantum dots (QDs) are very promising candidates for future
implementation of quantum computations: an electron spin in a QD
is a natural two-state quantum system, which can be efficiently
manipulated by the external magnetic fields and gate voltages
\cite{dqde05,Nielsen02}. Also, the QD-based architectures are
potentially scalable, and rely on well-developed semiconductor
technology \cite{Loss98}. However, due to interaction with the
environment, the electron spin loses coherence very quickly, on a
time scale of order of nanoseconds for typical GaAs QDs
\cite{dqde05}. It is vitally important for realization of QD-based
quantum computing to understand the decoherence dynamics in
detail, in order to find practical ways of decoherence
suppression. Moreover, decoherence of open systems is of
fundamental interest for understanding of the quantum phenomena
taking place in mesoscopic systems
\cite{Leggett87,Gardiner00,Yu06}, and therefore attracts much
attention from scientists working in the areas of nanoscience,
spintronics, and quantum control.

Among different sources of decoherence relevant for an electron
spin in a QD, the decoherence by the bath of nuclear spins (spin
bath) is dominant for magnetic fields less than a few Tesla, and
experimentally relevant temperatures of tens or hundreds of
milliKelvin. Much research has been focused on the case of a large
external magnetic field or large bath polarizations, where the
perturbation theory allows an extensive analysis
\cite{Coish04,Schliemann03, Khaetskii02, Deng06}. But the
interesting and experimentally relevant regime of moderate
magnetic fields and/or moderate bath polarization has received
much less attention.

Below, we study in detail the influence of moderate magnetic
fields and bath polarization on the longitudinal decoherence of a
single spin in a quantum dot, and two spins located in neighboring
QDs, where the perturbation theory is not applicable. In this
regime, as the magnetic field or the bath polarization increase,
the dynamics of the electron spins undergoes a transition from
simple overdamped decay to underdamped oscillations, and these
oscillations can be detected using existing experimental schemes.
In this transition, the increase of bath polarization affects the
decoherence dynamics in exactly the same manner as the increase of
the magnetic field.
Our results also show that suppression of decoherence requires
very large bath polarizations, suggesting that the use of such
methods as dynamical decoupling \cite{DD,echo05,dqde05}, or
control of the nuclear spins distribution \cite{Stepanenko06}, may
be more practical.

Theoretical description of the spin-bath decoherence is a very
complex problem \cite{ProkStamp}, where strong correlation and
essentially non-Markovian bath dynamics play an important role
\cite{Dobrovitski03,Coish04,Lages05}. In contrast with the well
studied boson-bath decoherence \cite{Leggett87,Gardiner00},
decoherence by a spin bath has not yet been understood in detail,
especially in moderate external magnetic fields.

\section{Single electron spin in a QD}

An electron spin in a QD interacting with a bath of nuclear spins
is described by the Hamiltonian which includes the Zeeman energy
of the electron spin in the external magnetic field $B_0$ and the
contact hyperfine coupling \cite{Al-Hassanieh06, Dobrovitski06}:
${\cal H} = H_0 S_z + \sum_{k=1}^N A_k {\bf S} \cdot {\bf I}_k$,
where ${\mathbf S}$ is the operator of the electron spin,
${\mathbf I}_k$ is the operator of the $k$-th bath spin
($k=1,2,\dots,N$), and $A_k=(8\pi/3)g_e^*\mu_Bg_n\mu_n u({\mathbf
x}_k)$ is the contact hyperfine coupling which is determined by
the electron density $u({\mathbf x}_k)$ at the site ${\mathbf
x}_k$ of the $k$-th nuclear spin and by the Land\'e factors of the
electron $g_e^*$ and of the nuclei $g_n$. The terms omitted in the
above Hamiltonian, such as the Zeeman energy of the nuclear spins,
the anisotropic part of the hyperfine coupling, etc., are small,
and can be safely neglected at the nanosecond timescale, which is
considered here.

In spite of apparent simplicity of the Hamiltonian, it is very
difficult to determine the dynamics of the central spin ${\bf
S}(t)$. Previously, the quasi-static approximation (QSA) for the
nuclear spin bath, which treats $\sum_k A_k {\bf I}_k$ as a
constant, has often been invoked
\cite{Merkulov02,Semenov03,Erlingsson04,Taylor06}. Analytical
calculations beyond the QSA have been carried out only for large
magnetic field and/or large bath polarizations
\cite{Coish04,Khaetskii02,Schliemann03, Deng06}, where the
perturbative approach is valid. Although many qualitative
arguments support QSA, its validity has not yet been checked in
detail. Below, along with explicit analytical solutions, we
provide direct verification of our analytics by employing the
exact numerical simulations for medium-size ($N=20$) baths, and
approximate simulations for large ($N=2000$) baths. The agreement
between all three methods ensures us that our findings do not
depend on the approximations involved.

The analytical calculations within the QSA can be performed in
different ways which all give the same results. The conceptually
simplest way is to assume a uniform electron density in the QD, so
that all $A_k$ are the same, $A_k=A$, and the Hamiltonian can be
written as ${\cal H} = H_0 S_z + A\bf S \cdot \bf I$,
where ${\bf I}=\sum_k {\bf I}_k$ is the total nuclear spin of the
bath. This Hamiltonian can be analyzed exactly, since ${\bf I}^2$
and $I_z+S_z$ are the integrals of motion.

\begin{figure}
\includegraphics[width=3.25in]{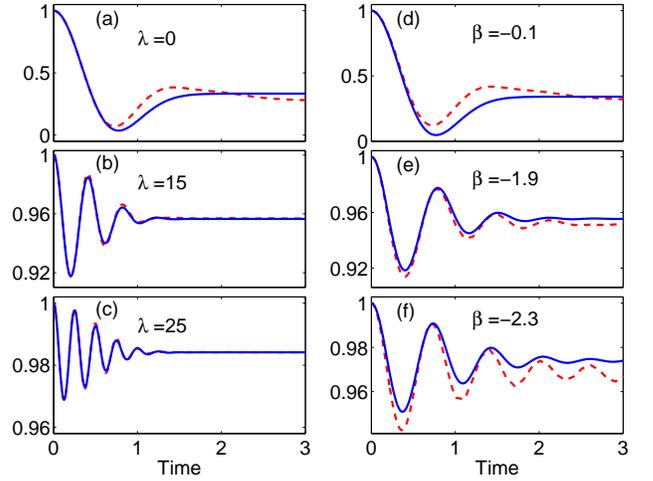}
\caption{(Color online) Electron spin decoherence in various
magnetic fields [(a), (b), (c)] and with polarized initial nuclear
spin baths [(d), (e), (f)] for $N=20$. The red dashed curves
denote the exact numerical simulations results for $N=20$, and the
blue solid curves correspond to the analytical results. The
numerical results agree well with the analytical predictions, and
the underdamped oscillations appear once $\lambda$ or $\kappa$ is
larger than $\sqrt N$.}
\label{fig:rbp}   
\end{figure}

\begin{figure}
\includegraphics[width=3.25in]{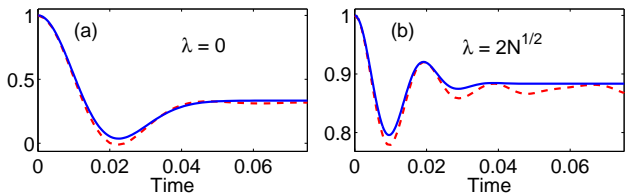}
\caption{(Color online) The same as Fig. \ref{fig:rbp} but for
$N=2000$ nuclear bath spins in zero magnetic field (a) and in
$\lambda = 2\sqrt N$ (b) with initially unpolarized bath.}
\label{fig:rb2}   
\end{figure}

First, we consider an unpolarized bath, with the density matrix
$\rho_b(0) = (1/2^N) {\bf 1}_1\otimes\cdots\otimes{\bf 1}_N$,
where ${\bf 1}_k$ is the unity matrix for $k$-th bath spin. This
density matrix can be re-written in the eigenbasis of ${\bf I}^2$,
$I_z$ as $\rho_b(0)=\sum_{I=0}^{N/2}\sum_{M=-I}^I P(I,M)
|I,M\rangle \langle I,M|$ where $M=I_z$. Assuming that the initial
state of the electron is $|\uparrow\rangle$, the
quantum-mechanical average of the $z$-component of the electron
spin at time $t$ is
\begin{equation}
\sigma_z(t) \equiv 2\langle S_z(t)\rangle =
    \sum_{I,M} P(I,M) {B^2
    + C^2\cos\Omega t\over \Omega^2} \label{eq:rsum}
\end{equation}
where $C=A\sqrt{(I-M)(I+M+1)}$, $B=H_0+A(M+1/2)$, and $\Omega^2 =
B^2+C^2$. Due to the symmetry of the problem and the initial
condition, $\sigma_x(t)=\sigma_y(t)=0$, so we omit these
components. The distribution function $P(I,M)$ can be calculated
\cite{Melikidze04}
and for large $N$ and $I$ it is approximated by a Gaussian distribution
$P(I,M) \approx (I /D\sqrt{2\pi
D}\;)e^{-I^2/2D}$ where $D=N/4$. Replacing the summation by
integration in Eq.~(\ref{eq:rsum}), we find
\begin{equation}
\label{eq:rb}
\sigma_z (t) = 1 - 2 W(\lambda,D;t),
\end{equation}
and the function $W(\lambda,D;t)$ has the form
\begin{eqnarray}
W &= & {D\over \lambda^2}-{D \over \lambda^2} e^{-Dt^2/2}
    \cos\lambda t
     +i \sqrt{\pi\over 2}{D^{3/2}\over 2\lambda^3} e^{-\lambda^2/2D}
     \nonumber \\
    &\times & \left[\Phi({Dt-i\lambda\over \sqrt{2D}}) - \Phi({Dt+i\lambda\over \sqrt{2D}})
    +2\Phi({i\lambda \over\sqrt{2D}})\right]
    \label{eq:wfunc}
\end{eqnarray}
where $\Phi(x)$ is the error function. In this equation, we took
$A=1$, and introduced the notation $\lambda=H_0/A$; this
corresponds to normalized energy and time scales, so that the time
$t$ is measured in the units of $1/A$. The dynamics of
$\sigma_z(t)$ for several values of $\lambda$ are shown in
Figs.~\ref{fig:rbp} and \ref{fig:rb2}.

Before discussing the above results, we consider next a polarized
bath, assuming that its initial state is described by a density
matrix $\rho_b(0) = (1/Z_\beta)\exp(-\beta M)$, where $\beta$ is
the inverse spin temperature, and $Z_\beta = [2\cosh(\beta/2)]^N$
is statistical sum,
and the initial
bath polarization is $p \equiv
\langle M\rangle / (N/2) = -\tanh(\beta/2)$. For large spin
temperatures, the polarization is small, $p\approx -\beta/2$,
and we approximate
$P(I,M) \approx  (I/ D\sqrt{2\pi D}\;) e^{-D\beta^2/2} e^{-I^2/2D}
e^{-\beta M}$, so that
\begin{equation}
\label{eq:rps}
\sigma_z (t) = 1 - 2 W(\kappa,D;t),
\end{equation}
where $\kappa = D \beta$, and the function $W(\kappa,D;t)$ is
defined by Eq.~(\ref{eq:wfunc}). The dynamics of $\sigma_z(t)$ for
several values of $\beta$ is shown in Fig.~\ref{fig:rbp}.

The functional form of Eqs.~(\ref{eq:rb}) and (\ref{eq:rps}) is
identical, up to replacing $\kappa$ by $\lambda$, so the small
nonzero bath polarization $p$ affects the central spin in exactly
the same way as the external field of the magnitude $H_0=-pAN/2$,
equal to the average Overhauser field exerted on the central spin
by the nuclear bath.
Indeed, the noticeable average magnetization $\langle
M\rangle=pN/2$ of the polarized bath leads to a noticeable average
Overhauser field, but the variation of the magnetization
$\langle(\Delta M)^2\rangle=(1-p^2)N/4$ changes very little at
small $p$, and, correspondingly, the spread in the Overhauser
fields is almost unchanged.

From Eqs.~(\ref{eq:rb}) and (\ref{eq:rps}), we see that
$\sigma_z(t)$ always decays with characteristic time
$T_2^*=\sqrt{8/(NA^2)}$. E.g., for $H_0=0$ (or $p=0$), we
reproduce a well-known result $\sigma_z(t) = (1/3) +
(2/3)(1-Dt^2)\exp(-Dt^2/2)$, i.e.\ $\sigma_z(t)$ first falls to
the value $\approx 0.036$, then increases and saturates at $1/3$.
However, for $H_0\neq 0$ (or $p\neq 0$), due to the oscillatory
terms in Eq.~(\ref{eq:wfunc}), $\sigma_z(t)$ can exhibit
oscillations, provided that $\lambda$ (or $\kappa$) is comparable
or larger than $\sqrt{D}\sim\sqrt{N}$. This transition, from
smooth decay at small field/polarization, to the oscillations at
larger field/polarization, is similar to the well-known transition
in the dynamics of a damped oscillator: the evolution of the
central spin is overdamped (or underdamped) depending on whether
the decay time $T_2^*$ is larger (or smaller) than the ``bare''
oscillation frequency determined by $\lambda$ or $\kappa$. Note
that the crossover value for magnetic field (polarization) is of
order of $\sqrt{N}$, beyond the range of applicability of the
perturbation theory \cite{Coish04,Deng06}.

For a typical GaAs QD with the electron delocalized over $N=10^6$
nuclear spins, $A\sim A_0/N\sim 10^{-4}$ $\mu$eV (where
$A_0\approx 0.1$ meV is the hyperfine coupling for an electron
localized on a single nucleus \cite{Paget77}). The corresponding
decoherence time $T_2^* \sim 10$ ns (taking into account the
$I=3/2$ spins of $^{69}$Ga, $^{71}$Ga and $^{75}$As), as confirmed
by recent experiments \cite{dqde05,Taylor06}. To observe
oscillations for a single electron spin, a very modest external
field of order of 3 mT, or polarization of order of 0.5\% is
needed.

It is noteworthy that the decoherence time remains practically
constant in the course of transition from smooth to oscillatory
decay. The decoherence time is determined by the spread in the
Overhauser fields, which is proportional to the variation of the
bath magnetization $\langle(\Delta M)^2\rangle=(1-p^2)N/4$, and it
is little affected by small bath polarizations $p$ \cite{Coish04,
Cerletti05}. For a moderate bath polarization, the equivalence
between the external field and the nonzero bath polarization
disappears, and the analytical expression for $\sigma_z(t)$
becomes rather complex, involving hypergeometric functions
$_2F_1(a,b,c,z)$ of complex argument, so we do not present it
here. In this regime, the decoherence time does not increase much,
as shown in Fig. \ref{fig:rbp}(e) and (f).

In order to substantially increase the decoherence time, an
extremely large bath polarization is required, as suggested by the
results from the perturbation method \cite{Coish04,Schliemann03,
Khaetskii02, Deng06}. Such a large bath polarization is beyond the
scope of our paper. Currently, strongly polarized baths are
difficult to achieve experimentally, and such methods as narrowing
the nuclear spin distribution \cite{Stepanenko06}, dynamical
decoupling \cite{DD} and spin echo techniques
\cite{echo05,dqde05}, may be more practical for suppression of
decoherence.

The quasistatic bath approximation is far from reality. E.g., for
$H_0=0$ and $p=0$, the QSA predicts saturation of $\sigma_z(t)$ at
1/3 for $t\to\infty$, which is just an artifact of the
approximation: the detailed analysis shows that $\sigma_z$ slowly
decays (as $1/\ln{t}$) to zero
\cite{Erlingsson04,Coish04,Al-Hassanieh06}. However, we expect QSA
to be valid at times of order of $T_2^*$, when the bath's internal
dynamics is not yet important.

For verification, we perform exact numerical simulation with
medium-size baths of $N=20$ spins. A real QD is approximated by
taking nuclear spins located at the sites of the $4\times 5$ piece
of a square lattice, with the lattice constant $a=1$. Assuming a
parabolic confining potential, the electron density is
approximated as 2-D Gaussian with the widths of $w_x=w_y=1.5$, and
with the center shifted by $d_x=0.1$ and $d_y=0.29$ along the $x$-
and $y$-axis respectively. For comparison with the analytical
results above, the effective coupling constant is defined as
$A=\left(\sum_k A_k^2/N\right)^{1/2}$. The decoherence dynamics is
simulated by directly solving the time dependent Schr\"odinger
equation for the wave function of the full many-spin system
(central spin plus the bath), using the Chebyshev polynomial
expansion of the evolution operator as described in
Ref.~\onlinecite {Dobrovitski03}.

Fig.~\ref{fig:rbp}(a), (b), and (c) show analytical [obtained from
Eq.~(\ref{eq:rb})] and numerical results for $\sigma_z(t)$, for
different magnetic fields. The panels (d), (e), and (f) present
$\sigma_z(t)$ for polarized baths with different initial
polarizations.
For the analytical curve in panel (d) the small-polarization
formula Eq.~(\ref{eq:rps}) has been used, while the analytical
curves in panels (e) and (f) have been calculated from the formula
for moderate initial polarization. The agreement of the analytics
and the exact numerics is good. The difference is caused only by
the modest number $N=20$ of the bath spins used for numerical
simulations, but exact simulations even for $N=50$ are beyond the
capabilities of modern computers (since the computation time and
memory grow exponentially with $N$).

To study large baths with $N=2000$, we use the coherent state
P-representation described in Ref.~\onlinecite{Al-Hassanieh06}.
Although approximate, this numerical approach demonstrates
excellent accuracy. Figure~\ref{fig:rb2} presents the results for
$\sigma_z(t)$ obtained analytically, from Eq.~(\ref{eq:rb}), and
numerically, from P-representation simulations with $N=2000$
spins. For $N=2000$, the agreement between the analytics and the
numerics is very good. Overall, the data presented in
Figs.~\ref{fig:rbp} and \ref{fig:rb2} justify the use of QSA, so
that we expect our predictions to be valid for real QDs with
$10^6$ nuclear spins.

\begin{figure}
\includegraphics[width=3.25in]{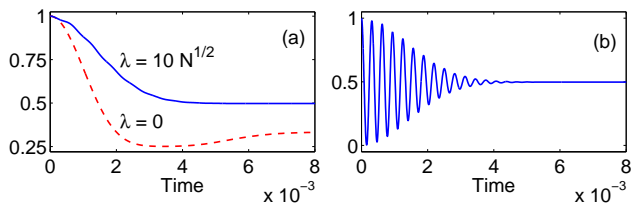}
\caption{(Color online) Time evolution of the probability of the
singlet state of the two electrons in a double QD by applying (a)
parallel magnetic fields $\lambda_1=\lambda_2=\lambda$ and (b)
antiparallel magnetic fields $\lambda_1=-\lambda_2=10\sqrt N$ with
$N_1=N_2=N=10^6$. Underdamped Rabi oscillations appear once the
difference in the magnetic fields applied onto each QD is large
enough.}
\label{fig:dqd}   
\end{figure}

\section{Two electron spins in double QD}

Measurement of the Rabi oscillations for a single electron spin in
a QD has not yet been achieved \cite{Koppens06}. Recent
experiments use two electron spins in two neighboring QDs
\cite{dqde05,Taylor06}. The spins are prepared in the singlet
state, and the measured quantity is the probability $P_S(t)$ to
stay in the singlet state after time $t$. The oscillations
described above can be also detected in this double-dot setup.
When the coupling between the two electron spins is negligible,
the Hamiltonian of the double-QD system is
${\cal H} = H_{01}
S_{1z} +  H_{02} S_{2z} + \sum_{j=1}^{N_1}A_j{\bf S}_1\cdot{\bf
I}_j + \sum_{k=1}^{N_2} A_k{\bf S}_2\cdot{\bf I}_{k}$
where indices 1 and 2 denote the quantities describing left and
right QD, respectively (e.g., $H_{01}$ and $H_{02}$ are the
magnetic fields acting on the left and right spins, respectively).
Note that the evolution of QD 1 and 2 is not independent because
of the initially entangled singlet state, although the Hamiltonian
is separable. Using the quasistatic approximation, we get
\begin{eqnarray}
\label{eq:dqd1}
P_S&=&{1\over 2} U(\lambda_1, D_1) U(\lambda_2, D_2)
    + {1\over 2} W(\lambda_1, D_1) W(\lambda_2, D_2) \nonumber \\
    &+&F(\lambda_1, D_1) F(\lambda_2, D_2) +
    G(\lambda_1, D_1) G(\lambda_2, D_2)
\end{eqnarray}
where the function $W(\lambda,D)$ is defined by
Eq.~(\ref{eq:wfunc}), and
$F(\lambda, D) = 1/2\left(1 +
    [\cos\lambda t - (Dt/\lambda)\sin\lambda t] e^{-Dt^2/2}\right)$,
$G(\lambda, D) = 1-F(\lambda, D)
    -W(\lambda,D)$,
$U(\lambda, D) = (1/ \lambda^2)
    [D\lambda t\cos\lambda t + (\lambda^2-D)\sin\lambda
    t] e^{-Dt^2/2}$.
Figure \ref{fig:dqd}(a) illustrates dynamics of $P_S(t)$ for
unpolarized baths, in the case of uniform magnetic field
$H_0=H_{01}=H_{02}$. $P_S(t)$ decays in the beginning, and
saturates at non-zero value at long times. The saturation value is
$1/3$ for zero field, and increases with the magnetic field,
reaching $1/2$ for strong fields \cite{Taylor06}.

However, if the difference between $H_{01}$ and $H_{02}$ is
comparable or larger than $A\sqrt{N_{1,2}}$ (which corresponds to
few milliTesla for realistic GaAs QD), the probability $P_S(t)$
exhibit oscillations [see Fig.~\ref{fig:dqd}(b)], analogous to the
oscillations of $\sigma_z(t)$ in the single-QD case above. In
experiments, a non-uniform magnetic field can be created e.g.\ by
micromagnets \cite{Wrobel04}. Another opportunity is using
non-uniformly polarized nuclear spin baths in the left and right
QDs. In analogy to the single-QD calculations, it can be shown
that $P_S(t)$ in the case of nonzero initial polarization is still
given by Eqs.~(\ref{eq:dqd1}), with replacement of $\lambda_{1,2}$
by $\kappa_{1,2}$. The difference in polarization should be about
0.5\% for the oscillations to appear.

\section{Conclusion}

In summary, we study in detail the influence of magnetic fields
and bath polarization on the decoherence of a single spin in a
quantum dot, and two spins located in neighboring QDs. We focus on
the regime of moderate fields and polarizations, where the
perturbation theory is not yet applicable, using both analytical
tools (the quasi-static bath approximation) and the numerical
simulations (exact, for medium-size baths with 20 spins, and
approximate, for large baths with 2000 spins). The agreement
between all three approaches is good, so we believe our results
are applicable to real QDs with millions of nuclear spins. The
nonzero bath polarization and the external magnetic field
influence the decoherence dynamics in exactly the same way, and
lead to a transition from smooth decay to oscillations once the
field (polarization) exceeds a certain crossover value. This
transition can be observed in experiments with a single QD, and
with two quantum dots. Our results show that substantial increase
of the decoherence time requires extremely large bath
polarizations, so that such methods as dynamical decoupling
\cite{DD,echo05,dqde05} or control of the nuclear spins
distribution \cite{Stepanenko06} may be more practical for
controlling decoherence of QD-based qubits.

\section{Acknowledgement}

This work was supported by the NSA and ARDA under Army Research
Office (ARO) contract DAAD 19-03-1-0132. This work was partially
carried out at the Ames Laboratory, which is operated for the U.
S. Department of Energy by Iowa State University under contract
No. W-7405-82 and was supported by the Director of the Office of
Science, Office of Basic Energy Research of the U. S. Department
of Energy. K.A.A. and E.D. are supported by the NSF grant No.
DMR-0454504.

\end{document}